\documentclass{new_tlp}
\usepackage{mathptmx}

\usepackage{verbatim}
\usepackage{mathptmx}
\usepackage{amssymb}
\usepackage{graphicx}
\usepackage{url}

\usepackage{times}
\usepackage{helvet}
\usepackage{courier}

\usepackage{epsfig}
\usepackage{color}
\usepackage{amssymb}
\usepackage{amstext}
\usepackage{amsmath}
\usepackage{multicol}

\newcommand{\fa}{\mathit{fa}}
\newcommand{\tr}{\mathit{tr}}
\newcommand{\good}{\mathit{good}}
\newcommand{\assign}{\mathit{assign}}
\newcommand{\inn}{\mathit{in}}
\newcommand{\false}{\mathit{false}}
\newcommand{\ok}{\mathit{ok}}
\newcommand{\next}{\mathit{next}}
\newcommand{\clsat}{\mathit{clsat}}

\newcommand{\Ma}{\mathit{mary}}

\newcommand{\To}{\mathit{tom}}
\newcommand{\current}{\mathit{current}}
\newcommand{\trained}{\mathit{trained}}
\newcommand{\visitor}{\mathit{visitor}}
\newcommand{\approved}{\mathit{approved}}
\newcommand{\access}{\mathit{unauthorized\_access}}
\newcommand{\breach}{\mathit{security\_breach}}
\newcommand{\account}{\mathit{account}}
\newcommand{\staff}{\mathit{staff}}

\newcommand{\Da}{\mathit{dan}}
\newcommand{\Wh}{\mathit{warehouse}}
\newcommand{\Warehouse}{\mathit{warehouse}}
\newcommand{\sem}{\mathit{sem}}
\newcommand{\nf}{\mathbf{\,not\;}}
\newcommand{\nop}[1]{}

\newcommand{\PP}{\mathcal{P}}       

\renewcommand{\H}{\mathcal{H}}
\newcommand{\Q}{\mathcal{Q}}

\newcommand{\Dom}{\mathcal{D}}
\newcommand{\T}{\mathcal{T}}       
\newcommand{\At}{\mathit{At}}
\newcommand{\Cl}{\mathit{Cl}}

\newcommand{\clause}{\mathit{clause}}
\newcommand{\pos}{\mathit{pos}}
\newcommand{\ngtd}{\mathit{ngtd}}
\newcommand{\chse}{\mathit{choose}}
\newcommand{\gate}{\mathit{gate}}
\newcommand{\true}{\mathit{true}}
\newcommand{\holds}{\mathit{holds}}
\newcommand{\sometrue}{\mathit{sometrue}}
\newcommand{\allfalse}{\mathit{allfalse}}
\newcommand{\sat}{\mathit{sat}}
\newcommand{\clfalse}{\mathit{clfalse}}
\newcommand{\bad}{\mathit{bad}}
\newcommand{\goal}{\mathit{goal}}

\newcommand{\A}{\mbox{$\cal A$}}

\newcommand{\IC}{\mathcal{C}}
\def\<{\mbox{$\langle$}}
\def\>{\mbox{$\rangle$}}

\newtheorem{example}{Example} 
\newtheorem{definition}{Definition} 
\newtheorem{proposition}{Proposition} 
\newtheorem{theorem}{Theorem} 

\newcommand\bcmdtab{\noindent\bgroup\tabcolsep=0pt%
  \begin{tabular}{@{}p{10pc}@{}p{20pc}@{}}}
\newcommand\ecmdtab{\end{tabular}\egroup}

\long\def\comment#1{}

  \title[A Measure of Arbitrariness in Abductive Explanations]
        {A Measure of Arbitrariness in Abductive Explanations}

  \author[Caroprese, Trubitsyna, Truszczy{\'n}ski, Zumpano]
         {Luciano Caroprese, Irina Trubitsyna\\
        DIMES, Universit\`a della  Calabria, Cosenza, Italy\\
         \email{{caroprese,irina}@deis.unical.it}
\and
Miros{\l}aw Truszczy{\'n}ski\\
Department of Computer Science, University of Kentucky, Lexington, USA\\
\email{mirek@cs.uky.edu}
\and
Ester Zumpano \\
         DIMES, Universit\`a della  Calabria, Cosenza, Italy\\
 \email{zumpano@deis.unical.it}\\
}

\jdate{March 2003} \pubyear{2003}
\pagerange{\pageref{firstpage}--\pageref{lastpage}}
\doi{S1471068401001193}

\vspace{-2mm}

\begin{document}

\label{firstpage}

\maketitle

\vspace{-2mm}
\begin{abstract}
\begin{quote}
We study the framework of abductive logic programming extended
with integrity constraints. For this framework, we introduce a new measure of
the simplicity of an explanation based on its degree of \emph{arbitrariness}:
the more arbitrary the explanation, the less appealing it is, with
explanations having no arbitrariness --- they are called \emph{constrained} 
--- being 
the preferred ones. In the
paper, we study basic properties of constrained explanations. For the case
when programs in abductive theories are stratified we establish results
providing a detailed picture of the complexity of the problem to decide
whether constrained explanations exist.
\end{quote}

\noindent
\textit{To appear in Theory and Practice of Logic Programming (TPLP).}
\end{abstract}
\vspace{-5mm}
\section{Introduction}
Abduction is an important form of nonmonotonic reasoning. The concept
was introduced in the late 19th century by the American philosopher
Charles Sander Peirce \shortcite{Peirce1955} as an inference scheme aimed at
deriving potential explanations of observations.\footnote{Peirce gave
abduction the following informal interpretation: ``The surprising fact,
$C$, is observed; but if $A$ were true, $C$ would be a matter of course:
hence, there is reason to suspect that $A$ is true.''}
A general characteristic of abductive reasoning is the existence of
multiple abductive explanations, which are typically not equally
compelling. Therefore, identifying a subclass, possibly narrow,
of ``preferred explanations'' is an important problem. Following
the Occam's principle, a typical approach is to identify as
``preferred'' those explanations that are, in some sense, \emph{simple}.
Several concepts of simplicity were considered in the literature,
most notably those based on minimality with respect to inclusion and
cardinality.
In the context of logic programming, abduction was first studied
by Eshghi and Kowalski \shortcite{Eshghi-Kowalski1989}, and then
by Kakas and Mancarella \shortcite{KakasM90a} under the \emph{brave
reasoning} variant of the stable-model semantics. That work established
\emph{abductive logic programming} as an important subarea of abduction,
where the background theory is represented by a logic program, often with
negation in the bodies and disjunction in the heads, under any of the
standard logic programming semantics \cite{KakasKT92,Denecker-Kakas2002}.
This paper is concerned with the problem of limiting the space of
explanations in the framework of abductive logic programming extended
by integrity constraints. We introduce a new measure of the quality
of an explanation in terms of its  \emph{arbitrariness} and propose to
consider as ``preferred'' only those explanations that minimize
arbitrariness. Our approach can be applied with any of the standard
semantics of logic programs. 

\begin{example}
Let us consider the following scenario. It is Saturday and Bob is
known to ski on Saturdays if it is not snowing. And when he is not skiing,
he is on campus. We can represent this information by the logic program:
\begin{center}
\{ $saturday.$ \ \ \  $skiing \leftarrow saturday, \nf \mathit{snows}.$  \ \ \
 $\mathit{on\_campus} \leftarrow \nf skiing.$
\}
\end{center}

If it's Saturday and we see Bob on campus, we can abduce it is snowing. Otherwise,
Bob would be skiing. To put it differently, given our background knowledge,
the fact $\mathit{snows}$ is an abductive explanation (explanations will be 
formally defined in the next section) to our observation 
$\mathit{on\_campus}$.~\hfill $\Box$
\end{example}

In this example, there is only one explanation of the observation
(assuming all that is relevant to our reasoning has been mentioned).
However, a general feature of abductive reasoning is the existence of
\emph{multiple} explanations. Typically, they are not all equally likely.
Thus, narrowing down the range of possible explanations to a smaller set
of ``most likely'' or ``preferred'' ones becomes an important problem.
The key to it is a well-motivated notion of ``preferred''. We will now
present an example meant to develop intuitions behind the notion of 
``preferredness''.

\begin{example}\label{Cont-Motivating-Example}
Let us consider the following scenario modeling security breaches in an
information system. A security breach at a component of the system may
only occur when a person with an account makes an unauthorized access.
Regular staff personnel have accounts on the system if they complete
training and have their security clearance current. Visitors may also
be granted an account but only with an approval by the head of the IT
department. This situation can be described by the following program:
\begin{align*}
\account(X) \leftarrow & \ \staff(X), \trained(X), \current(X).\\
\account(X) \leftarrow & \ \visitor(X), \approved(X).\\
\breach(W) \leftarrow & \ \access(W,X), \account(X).
\end{align*}
Let us also assume that $\To$ and $\Ma$ are regular staff members
and $\Da$ is a visitor (there may also be other individuals in these
groups and additional ones not mentioned as staff or visitors in the
program), and that the system has information that $\To$ completed training.
That is, the program also contains facts
\begin{align*}
&\staff(\To).\ \ \staff(\Ma).\ \ \visitor(\Da).\ \ \trained(\To).
\end{align*}
If we observe $\breach(\Warehouse)$ (the security of $\Warehouse$ was
compromised), there are several possible explanations. Below we
list some of them:

\vspace{-2mm}
\begin{align*}
E_\To &= \{\access(\Wh,\To),\current(\To)\}\\
E_\Ma &= \{\access(\Wh,\Ma),\trained(\Ma),\current(\Ma)\}\\
S_u   &= \{\access(\Wh,U),\staff(U),\trained(U),\current(U)\},\\
      &\quad\quad \mbox{where $U$ is a name in the domain, possibly
                  not mentioned in the program},
\end{align*}
\begin{align*}
E_\Da &= \{\access(\Wh,\Da),\approved(\Da)\}\\
V_u   &= \{\access(\Wh,U),\visitor(U),\approved(U)\},\\
      &\quad\quad \mbox{where $U$ is a name in the domain, possibly
                  not mentioned in the program}\\
E_{\To,\Da} &= \{\access(\Wh,\To),\current(\To),\\
            &\quad\quad   \access(\Wh,\Da), \approved(\Da)\}.
\end{align*}
The key question is whether there are principled ways to eliminate some of
these explanations as less likely than others. 
~\hfill $\Box$
\end{example}

Most approaches to the problem of selecting preferred explanations follow
the Occam's principle of parsimony that entities should not be multiplied
unnecessarily and that among possible explanations the simplest one tends
to be the right one. The first part of that principle is non-controversial.
However, simplicity is a notoriously complex concept and different
formalizations of it are possible! They range from the standard one based
on the subset minimality, to its versions and refinements that require
minimum cardinality, minimum weight, or minimality under prioritization of
individual hypotheses \cite{EiterGL97}. In our example, the explanations
$E_\To$, $E_\Ma$, $S_U$, $U\neq \To$, $E_\Da$, and $V_U$, $U\neq \Da$,
are subset minimal and so, \emph{preferred} under the subset minimality
criterion. The explanations $S_\To$, $V_\Da$, and $E_{\To,\Da}$ are not.
If we use a more restrictive criterion of minimum cardinality, the preferred
explanations are $E_\To$ and $E_\Da$. Let us assume that there are reasons
to view each of them as wrong ($\To$ and $\Da$ can conclusively demonstrate
they were not involved). Under the subset minimality criterion, we now
prefer explanations $E_\Ma$, $S_U$, $U\neq \To$, and $V_U$, $U\neq \Da$,
while under the minimum cardinality criterion $E_\Ma$ and $V_U$, $U\neq \Da$,
are preferred.

Let us look more carefully at these last two types of explanations. They
are specific in that they speak about concrete individuals. However, the
explanations of the latter type select an \emph{arbitrary} individual
in the domain (explicitly mentioned or not, as long as it is not $\Da$),
with no particular reason to choose one over another. In other words,
investigators of the security breach would have no reason to start their
investigation with any particular person and the person they would
consider first would be arbitrarily selected. On the other hand, the
explanation $E_\Da$ connects the structural information
present in the program and the knowledge provided by the observation
in a non-arbitrary (\emph{constrained}) way --- it ``invents'' no new
entities and makes no arbitrary assumptions. Arguably, it would be the
explanation pursued first by the investigators.

One might argue that the family of explanations $V_U$ simply points to
the \emph{existence} of a person who managed to pass as a visitor with
an approval for an account and made an unauthorized access to the IT
system, and that this family of explanations can be represented as a
single explanation involving existentially quantified formulas (along
the lines of an early work by Pople \shortcite{Pople73}). However,
such ``existential'' explanations while no longer making arbitrary
choices lack \emph{specificity}.

In this paper we develop these intuitions into a formal measure of
the quality of explanations based on how arbitrary they are. To this
end, we introduce the notion of the \emph{degree of arbitrariness}
of an explanation: the smaller that value, the less arbitrary the
explanation. We define \emph{constrained} explanations as those with 0
degrees of arbitrariness, and propose them as \emph{preferred}. Our most
significant technical results establish the computational complexity
of the problems to decide whether a given explanation is constrained,
and whether a constrained explanation exists. We study the two problems for
the case when programs modeling background knowledge in abductive theories
are stratified. This is a common case in abductive logic programming.
From the technical standpoint it has two interesting aspects. First,
the two main semantics of logic programs, the stable-model semantics
\cite{gl88} and the well-founded semantics \cite{vrs91}, as well as
several others (but not the supported-model semantics \cite{ms92}) 
collapse in this case. Second, when the background knowledge is modeled 
by a stratified program under the stable-model semantics, all general 
semantics-independent schemas to define aductive explanations yield the 
same result. We show that for abductive theories with stratified logic 
programs, the two computational problems mentioned above are
coNP-complete and $\Sigma_2^P$-complete, respectively. By considering
special classes of stratified programs, we establish the main sources of
the complexity of the problem.

The remainder of the paper is organized into five sections. First, we
recall basic concepts of abductive logic programming and discuss four
general schemata to define explanations. Next, we introduce the key
concepts for the paper: the degree of arbitrariness in an explanation, and
then arbitrary and constrained explanations. The following section presents
our main complexity results. Next, we discuss related work where, in
particular, we note that our results are also relevant in the area of
view update repairs. We conclude by summarizing our contributions
and pointing out problems for future work.

\section{Abductive Logic Programs}\label{ALP}

We consider a fixed vocabulary $\sigma$ consisting of relation and
constant symbols (no function symbols). We write $\Dom$ for the set of
constants in $\sigma$ and assume that $\Dom$ is an infinite countable
set (in some examples, for the sake of simplicity of presentation, we
take $\Dom$ to be finite). We write $\H$ for the set of predicate
symbols in $\sigma$. For $\Q\subseteq\H$, we define $\Q^{\Dom}$ to be
the set of all ground atoms whose predicate symbols are in $\Q$. In
particular, $\H^\Dom$ is the Herbrand base of $\sigma$.

By $S$ we denote a semantics of logic programs. For now we do not
commit to any particular semantics. As usual, we assume only that $S$
is given in terms of Herbrand interpretations of $\sigma$, that is,
subsets of $\H^\Dom$.
For a logic program $\PP$ (in the vocabulary $\sigma$), we denote by
$\sem_S(\PP)$ the set of Herbrand interpretations of $\sigma$ that are
models of $\PP$ according to the semantics $S$. Since $\sigma$ is fixed,
we omit it from the notation. The most common choice for $S$ is the 
stable-model semantics \cite{gl88}.

We now recall the concept of an abductive theory in the logic programming
setting.

\begin{definition}[{\sc Abductive Theory}]
An \emph{abductive theory} $\T$ over a vocabulary $\sigma$ is a
triple $\<\PP,\A,\IC\>$, where
\begin{itemize}
\item[--] $\PP$ is a finite logic program over $\sigma$
\item[--] $\A\subseteq\sigma$ is a finite set of predicate names
called the \emph{abducible predicates}
\item[--] $\IC$ is a finite set of first-order sentences over $\sigma$ called
\emph{integrity constraints},
\end{itemize}
and where every rule in $\PP$ with an abducible predicate in the head
is a ground fact. \hfill$\Box$
\end{definition}

Informally, the program $\PP$ and the integrity constraints $\IC$ provide
a model (description) of the problem domain. The program $\PP$ defines
non-abducible predicates (those not in $\A$) in terms of abducible
predicates (those in $\A$). The integrity constraints in $\IC$ impose
domain constraints on abducible and non-abducible predicates in the
language. According to the definition, information about abducible
predicates is given in terms of ground facts. They explicitly specify
the extensions of abducible predicates. We refer to ground atoms based
on abducible predicates as \emph{abducibles}.
An \emph{observation} is a set of ground facts based on \emph{non-abducible}
predicates. An observation may ``agree'' with the program $\PP$ and
the integrity constraints $\IC$. But if it does not, we assume that
this ``disagreement'' is caused by the incorrect information about
the properties modeled by the \emph{abducible} predicates. Abductive
reasoning consists of inferring updates to the set of abducibles in
the program (removal of some and inclusion of some new ones) so that
the updated program, the integrity constraints and the observation
``agree.'' Each update that yields an agreement constitutes a possible
\emph{explanation} of the observation.
Several notions of agreement have been proposed. They are
defined in terms of entailment and satisfiability. A program $\PP$ is
\emph{consistent in the semantics $S$} if $\sem_S(\PP) \not= \emptyset$.
A consistent logic program $\PP$ \emph{skeptically} entails a sentence
$\varphi$ (in the same language as $\PP$), written $\PP\models_S\varphi$,
if for every $M\in \sem_S(\PP)$, $M\models \varphi$. A consistent logic
program $\PP$ \emph{skeptically} entails a set $\Phi$ of sentences,
written $\PP\models_S\Phi$, if $\PP\models_S\varphi$, for every
$\varphi\in\Phi$.

We now present four concepts of ``agreement'', three of which have
received significant attention in the literature.

\begin{definition}[\textsc{Agreement A \cite{Denecker-Kakas2002}}]
\label{def1}
Let $\PP$ be a program and $\IC$ a set of integrity constraints.
An observation $O$ \emph{agrees} with $\PP$ and $\IC$ if 
(1) $sem_S(\PP) \neq \emptyset$ ($\PP$ is consistent);
(2) $\PP\models_{S} \IC$; and
(3) $\PP\models_{S} O$.
\hfill $\Box$
\end{definition}

To get this notion of agreement, it is necessary that the program
be consistent and skeptically entail the observation. But, in addition,
the program must also satisfy the integrity constraints in a very strong
sense. Namely, all its models must satisfy them.

\begin{definition}[\textsc{Agreement B \cite{Denecker-Kakas2002}}]
\label{def2}
Let $\PP$ be a program and $\IC$ a set of integrity constraints.
An observation $O$ \emph{agrees} with $\PP$ and $\IC$ if 
(1) $\PP\models_{S} O$; and
(2) there is $M\in \sem_S(\PP)$ such that $M \models \IC$.
\hfill $\Box$
\end{definition}

This definition is a relaxation of the previous one. To get this
notion of agreement, at least one model of the program must satisfy
the integrity constraints (not all, as before, and here is where
the conditions are relaxed). This, in particular, means that the
program is consistent. Moreover, exactly as before, the program
must skeptically entail the observation.

\begin{definition}[\textsc{Agreement C}]
\label{def3}
Let $\PP$ be a program and $\IC$ a set of integrity constraints.
An observation $O$ \emph{agrees} with $\PP$ and $\IC$ if 
for every $M\in sem_S(\PP)$ such that $M\models\IC$, $M\models O$.
\hfill $\Box$
\end{definition}

Here to have an agreement, as in Definition \ref{def2}, we require that
at least one model of the program satisfies the integrity constraints and
we weaken the other condition. Namely, we now only require that the
observation holds in every model of the program satisfying the integrity
constraints and not in every model of the programs.

\begin{definition}[\textsc{Agreement D \cite{Baral-Gelfond1994}}]
\label{def4}
Let $\PP$ be a program and $\IC$ a set of integrity constraints.
An observation $O$ \emph{agrees} with $\PP$ and $\IC$ if there is $M\in
\sem_S(\PP)$ such that $M \models \IC$ and $M\models O$.
\hfill $\Box$
\end{definition}

This is the final relaxations of the conditions for an agreement.
There must be models of a program that satisfy the integrity constraints
(just as in Definition \ref{def3}). But it is enough that one of such
models (not all, as in Definition \ref{def3}) satisfies the observation.
Each of these types of agreement yields the corresponding notion of an
explanation.

\begin{definition}[\textsc{Abductive Explanation}]
\label{ae}
Let $\T=\<\PP,\A,\IC\>$ be an abductive theory and $O$ an observation.
A pair $\Delta=(E,F)$, where $E$ and $F$ are disjoint finite sets of abducibles
and $F\subseteq \PP$, is an \emph{explanation} (of type $A$, $B$, $C$, and
$D$) if $O$ is in agreement (of type $A$, $B$, $C$, and $D$, respectively)
with $\PP^\Delta= (\PP\cup E)\setminus F$ and $\IC$.
We denote the sets
of explanations of each type by $\Psi_{A} (O,\T)$, $\Psi_{B} (O,\T)$, $\Psi_{C} (O,\T)$ and
$\Psi_{D} (O,\T)$.  \hfill $\Box$
\end{definition}

The observations made above about each next concept of an agreement being
less restrictive, imply the following relationships between the corresponding
notions of explanations.

\begin{proposition}
For every abductive theory $\T$ and an observation $O$,
\small
$$ \Psi_{A} (O,\T)\subseteq \Psi_{B} (O,\T) \subseteq
\Psi_{C} (O,\T) \subseteq \Psi_{D} (O,\T).$$

\vspace*{-0.27in}
\hfill$\Box$
\normalsize
\end{proposition}

In general, the four concepts of explanations do not coincide. Which of
them to choose may depend on a particular application domain. This issue
is not of our concern here.
Instead, we focus on the key challenge of abductive reasoning to find
general principles that can narrow down a class of explanations and that
are independent of
what notion of an explanation is used. This is indeed an important problem
as even in the case of explanations of type $A$ multiple explanations are
possible and not all of them are equally compelling.

The two most commonly used principles are
subset minimality and cardinality minimality.
We say that an explanation $\Delta=(E,F)$ is \emph{subset minimal} if
there is no \emph{other} explanation $\Delta'=(E',F')$ such that
$E'\subseteq E$ and $F'\subseteq F$. Similarly, $\Delta=(E,F)$ is
\emph{cardinality minimal} if there is no explanation $\Delta'=(E',F')$
such that $|E'|+|F'| < |E|+|F|$. 
We propose in the next section yet another general principle based on the degree
of arbitrariness.

\section{How arbitrary is a solution?}\label{degree_of_freedom}

In this section we introduce the degree of \emph{arbitrariness} in
an explanation of an observation as a measure of how \emph{arbitrary}
the explanation is. Concepts introduced below apply to each of the four
basic models of abductive reasoning,
therefore we will use the generic term \emph{explanation} without denoting its specific type.
We start with examples to motivate our discussion.

\begin{example}\label{Cont2-Motivating-Example}
Let us consider the abduction problem presented in
Example~\ref{Cont-Motivating-Example}, and the explanations
$\Delta_{u}=(S_u,\emptyset)$ and $\Delta_{\Da}=(E_{\Da},\emptyset)$.
In each explanation $\Delta_u$, the constant $u$ can be replaced with
any other constant $u'$ in the vocabulary and the result, $\Delta_{u'}$,
is also an explanation. That is, the occurrences of $u$ are not
\emph{constrained} by the program or, to put it differently, they are
\emph{arbitrary}. In contrast, replacing the constant $\Da$ in
$\Delta_{\Da}$ with any other constant does not yield an explanation
(assuming there are no other visitors in the program but $\Da$). Thus,
there is no \emph{arbitrariness} in $\Delta_{\Da}$. In other words,
$\Delta_{\Da}$ is \emph{constrained} by the available
information.~\hfill$\Box$
\end{example}

We will use the idea of ``replaceability'' to formalize the notions of
arbitrariness and constrainedness, and their generalization, the degree
of arbitrarines.

\begin{definition}[\textsc{Occurrence}]
Let $p(x)$ be a ground atom, where $p$ has arity $n$ and $k$ is an integer
such that $1\leq k\leq n$. We denote by $p(x)[k]$ the constant in the position
$k$ in $p(x)$.

\noindent
If $E$ is a set of ground atoms, an \emph{occurrence} of a constant $c$ in $E$
is an expression of the form $p(x)^k$, where $p(x)$ is an atom in $E$,
and $p(x)[k]=c$.
\hfill $\Box$
\end{definition}

\begin{definition}[\textsc{Replacement Function}]
Let $\T=\<\PP,\A,\IC\>$ be an abductive theory, $E$ a set of abducibles
 (that is, $E\subseteq \A^{\Dom}$), and let $c$ be a constant occurring
in $E$. A \emph{replacement function} for $E$ and $c$ determined by a
\emph{non-empty} set $C$ of some (not necessarily all) occurrences of $c$
in $E$, is a function
$f_{E,C}: \Dom \rightarrow 2^{\mathcal{A^{\Dom}}}$
such that for each $x\in\Dom$, $f_{E,C}(x)$ is the set $E'$ obtained by
replacing with $x$ each constant $c$ in $E$ referred to by an occurrence 
in $C$.
\hfill
\hfill $\Box$
\end{definition}

We observe that, given a set $E$ and a constant $c$ occurring in $E$ in
$n$ places, the number of possible replacement functions is $2^n-1$.

\begin{example}
\label{ex4}
Let us consider the set $E=\{p(1,2),s(2,3)\}$ and the constant $2$.
The possible replacement functions are: (1)
$f_{E,C_1}$, where $C_1=\{p(1,2)^2\}$;
(2) $f_{E,C_2}$,  where $C_2=\{s(2,3)^1\}$; and (3)
$f_{E,C_3}$, where $C_3=\{p(1,2)^2,s(2,3)^1\}$.\hfill
\hfill $\Box$
\end{example}

\begin{definition}[\textsc{Independence of replacement functions}]
Let $f_{E,C_1}$ and $f_{E,C_2}$ be replacement functions for a set $E\subseteq
\A^\Dom$ and for constants $c_1$ and $c_2$, respectively.
We say that $f_{E,C_1}$ and $f_{E,C_2}$ are
\emph{independent} if $c_1\not= c_2$ or if $C_1\cap C_2=\emptyset$.~\hfill $\Box$
\end{definition}

Let us consider the replacement functions presented in Example \ref{ex4}.
While the functions $f_{E,C_1}$ and $f_{E,C_2}$ are independent,
the functions $f_{E,C_3}$ and $f_{E,C_1}$ (similarly, $f_{E,C_3}$ and
$f_{E,C_2}$) are not independent.

\begin{definition}[\textsc{Degree of Arbitrariness}]
\label{d-degree}
Let $\T=\<\PP,\A,\IC\>$ be an abductive theory, $O$ an observation,
$\Delta=(E,F)$ an explanation for $O$ wrt $\T$, and let $\xi$ be an
arbitrary constant in $\Dom$ not occurring in $E \cup O \cup \PP$. The
\emph{degree of arbitrariness} of $\Delta$, denoted $\delta(\Delta)$,
is the maximum number of \emph{pairwise independent} replacement functions
$f_{E,C}$ (not necessarily all for the same constant) such that
$\Delta'=(f_{E,C}(\xi),F)$ is an explanation for
$O$ wrt $\T$.~\hfill $\Box$
\end{definition}

Since the domain $\Dom$ is infinite, one can always find a constant $\xi$
not occurring in $E \cup O \cup \PP$. Moreover, since $E$ is finite, there
is an upper bound on the number of pairwise independent replacement functions
one can have for $E$. Finally, the specific choice of the replacement constant
$\xi$ does not affect the maximum. Thus, the degree of arbitrariness is well
defined. The examples below illustrate the concepts we have introduced above.

\begin{example}
\label{ex6}
Let $\T=\<\PP,\A,\emptyset\>$, where the program $\PP$ consists of the facts
$\{p(1),p(2),$ $q(1),q(2),q(3)\}$ and of a single rule $t \leftarrow p(X),
not\ q(X)$. Let us suppose that $p$ and $q$ are abducible predicates and
that $O=\{t\}$. The following pairs of sets of abducibles form explanations
for $O$ wrt $\T$: \\
$\Delta_{1} = (\emptyset,\{q(1)\}). \ $
$\Delta_{2} = (\emptyset,\{q(2)\}).\ $
$\Delta_{3} = (\{p(3)\},\{q(3)\}). \ $
$\Delta_{x} =  (\{p(x)\},\emptyset),\ \mbox{where $x\notin \{1,2,3\}$}$.

It is evident that $\delta(\Delta_{1})=\delta(\Delta_{2})=0$.
Similarly, it is evident that $\delta(\Delta_{3})=\delta(\Delta_{x})
=1$ (the only constant in the ``add'' part of these explanations can
be replaced with a fresh constant $\xi$ and the result is an explanation).
\hfill $\Box$
\end{example}

In Example \ref{ex6}, the explanation $\Delta_3$ is not satisfactory. Once
we decide to remove $q(3)$, there is no reason why we have to add $p(3)$.
Adding any atom $p(\xi)$, with $\xi\notin\{1,2\}$ works equally well. Thus,
the choice of the constant $3$ in $p(3)$ is arbitrary and not grounded in the
information available in the theory. Similarly, $\Delta_x$, where $x\notin
\{1,2,3\}$ is not satisfactory either. Here too, the choice of $x$ is not
grounded in the abductive theory and the observation. The explanations
$\Delta_1$ and $\Delta_2$ do not show this arbitrariness.

\begin{example}
\label{ex7}
Let $\T=\<\PP,\A,\emptyset\>$, where $\A=\{q,r,t\}$ and $\PP$ consists of
the facts $q(a,b)$, $q(a,c)$, and $r(a,b,c)$, and two rules
$p(X) \leftarrow q(X,Y), s(X,Y,Z)$, and $s(X,Y,Z) \leftarrow  r(X,Y,Z), t(X,Z)$.
Let us suppose $O=\{p(a)\}$. One can check that each of the following
pairs of sets of abducibles is an explanation:
\[
\begin{array}{lll}
\Delta_{x_1,x_2} & = & (\{q(a,x_1),r(a,x_1,x_2),t(a,x_2)\},\emptyset), \mbox{ where } x_1\neq c \mbox{, } x_1\neq b \mbox{ and } x_2\neq c\\
\Delta_{x_3}     & = & (\{r(a,b,x_3),t(a,x_3)\},\emptyset),
             \mbox{ where } x_3\neq c\\
\Delta           & = &(\{t(a,c)\},\emptyset).
\end{array}
\]
It is evident that if $x_1\not=x_2$, then $\delta(\Delta_{x_1,x_2})=2$.
Indeed changing all occurrences of $x_1$ or all occurrences of $x_2$
to a new constant $\xi$ results in an explanation. In addition, the
corresponding replacement functions for each constant and all its
occurrences are obviously independent (they concern different constants).
Finally, replacing either constant in only one position does not yield
an explanation. More interestingly, if $x_1=x_2=x$, $\delta(\Delta_{x,x})=2$,
too. Here, $x$ has four occurrences in $\Delta_{x,x}$: $q(a,x)^2$,
$r(a,x,x)^2$, $r(a,x,x)^3$ and $t(a,x)^2$. Let $f_1$ and $f_2$ be the
replacement functions for $x$ determined by the first two and the last two
positions, respectively. Then $f_1$ and $f_2$ are independent, and both
$(f_1(\xi),\emptyset)$ and $(f_2(\xi),\emptyset)$ are explanations. However,
three mutually
independent functions with this property cannot be found.
Similarly, one can see that $\delta(\Delta_{x_3})=1$ (all occurrences of
$x_3$ are free for a simultaneous change) and $\delta(\Delta)=0$ (neither
$a$ nor $c$ can be changed to a fresh constant). ~\hfill $\Box$
\end{example}

In this example, all explanations are minimal and so, the subset-minimality
criterion is not discriminating enough. However, $\Delta$ arguably is more
compelling than the other ones. It uses no extraneous constants, and all
constants occurring in it are constrained by the theory and an observation.
The lowest degree of arbitrariness criterion correctly identifies $\Delta$
as the only preferred explanation. We will refer to it as the
principle of \emph{minimum arbitrariness}.

The most compelling are those explanations
that have no arbitrariness at all. This suggests the notion of
constrained explanations.

\begin{definition}[\textsc{Constrained and Arbitrary Explanations}]
\label{constrained explanation}
Let $\T$ be an abductive theory $\<\PP,\A,\IC\>$, $O$ an observation,
and $\Delta$ an explanation for $O$ wrt $\T$. We say that $\Delta$ is
\emph{constrained} if $\delta(\Delta)=0$. Otherwise, $\Delta$ is
\emph{arbitrary}. ~\hfill $\Box$
\end{definition}

In the remainder of this paper we discuss the principle of the lowest
degree of arbitrariness, focusing our study primarily on the class of
constrained explanations. We start with some general observations.

The principle of minimum arbitrariness can be used with all types of
explanations we discussed in the previous sections. Moreover, it is
``orthogonal'' to other criteria one might consider when selecting
preferred explanations such as the subset or cardinality minimality.
Therefore, it can be combined with them. For instance, we might consider
as preferred those subset-minimal explanations that have the smallest
degree of arbitrariness (if we believe, that subset minimality is more
important than minimum arbitrariness). For instance, coming back to
Example \ref{Cont-Motivating-Example}, we see that the explanations
$(E_\To,\emptyset)$, $(E_\Ma,\emptyset)$, $(S_u,
\emptyset)$, $u\neq \To$, $(E_\Da,\emptyset)$, and $(V_u,\emptyset)$,
$u\neq \Da$, are subset minimal. Selecting from among them only the
constrained ones yields $(E_\To,\emptyset)$, $(E_\Ma,\emptyset)$, and
$(E_\Da,\emptyset)$.
Alternatively, we could take as preferred those explanations with the 
minimum degree of arbitrariness that are subset minimal (if we
believe that minimum arbitrariness should be the primary consideration).
In Example \ref{Cont-Motivating-Example}, the explanation $(E_{\To,\Da},
\emptyset)$ is constrained but not minimal. In general, the two concepts
are different. Similar examples can be provided for the minimum cardinality
criterion. We do not consider these and other possible combinations of
the principles in this paper but focus exclusively on the properties
of the principle of minimum arbitrariness.

The degree of arbitrariness of an explanation $(E,F)$ depends
only on the ``add'' part $E$; the ``delete'' component, $F$, has no
effect on arbitrariness. Intuitively, the reason is that we can delete
only those atoms that are in $\PP$. Thus, if we replace a constant
in an atom $p$ in $F$ with a fresh constant $\xi$, the effect simply
is that $p$ is no longer deleted. The same effect can be achieved by
considering $F\setminus\{p\}$ in place of $F$. It is natural to impose
on $F$ some requirements, such as subset or cardinality minimality.
However, as we noted above, we do not pursue this possibility here.

The next result shows that constrained explanations use only constants
that are mentioned in an abductive theory and an observation. This
property is consistent with the general principle of parsimony (Occam's
razor). It is important as it allows us to restrict the scope of searches 
for constrained explanations.\footnote{The proofs of all results we give
in the paper can be found in the appendix}

\begin{theorem}
\label{relevant}
Let $\T=\<\PP,\A,\IC\>$ be an abductive theory and $(E,F)$ a constrained
explanation of an observation $O$. Then every constant symbol occurring
in $(E,F)$ occurs in $\T$ or in $O$.
\end{theorem}

Finally, we note that the minimum arbitrariness criterion does
not impose any conditions on abducible predicates of arity 0
and some other criteria should be considered. Therefore, we are
primarily interested in the case when every abducible predicate
symbol has a positive arity.

\section{Computational Complexity}

Our primary technical contribution concerns reasoning about constrained
explanations. We are interested in the following two problems: deciding
whether a given explanation is constrained; and deciding whether a
constrained explanation exists.
We restrict attention to abductive theories with stratified programs and
assume that these programs are interpreted by the stable-model semantics. 
This is an important class of abductive theories. First, stratified programs 
are regarded as semantically ``non-controversial.'' Indeed, the stable-model 
semantics and the well-founded semantics coincide on stratified programs
and are generally accepted as providing them the correct meaning. Second, under 
the stable-model semantics, for abductive theories 
with stratified programs, the distinctions between the four types of 
explanations 
we introduced disappear. Formally, we have the following result.

\begin{theorem}
\label{same}
For every abductive theory $\T=\<\PP,\A,\IC\>$, where $\PP$ is stratified and
interpreted under the stable-model semantics,
and every observation $O$,
\small
$$ \Psi_{A} (O,\T)= \Psi_{B} (O,\T) =
\Psi_{C} (O,\T) = \Psi_{D} (O,\T).$$

\vspace*{-0.27in}
\hfill$\Box$
\normalsize
\end{theorem}
Because of Theorem \ref{same}, below we use the generic term \emph{explanation}
without specifying its type.

Let $\<\PP,\A,\IC\>$ be an abductive logic theory. We represent $\PP$
as the union of a set $B$ of all abducibles in $\PP$ and the set $R$ of 
the remaining facts and rules. We will consider the complexity of the problems
stated above under the assumption that $R$ and $\IC$ are fixed and input 
consists of $B$ and an observation $O$. 
We start our study of the complexity by noting the following two upper bounds.

\begin{theorem}\label{thm-upper}
Let $\A$ be a set of abducible predicates, $R$ a (fixed) stratified
program with no abducible predicates in the heads of its rules, and
$\IC$ a (fixed) set of integrity constraints.
\begin{enumerate}
\item The following problem is in coNP: given a set $B$ of abducibles,
an observation $O$, and a pair $\Delta=(E,F)$ of sets of abducibles,
decide whether $\Delta$ is a constrained explanation for $O$ wrt $\<R\cup B,\A,\IC\>$.
\item The following problem is in $\Sigma_2^P$: given a set $B$ of 
abducibles and an observation $O$, decide whether a constrained 
explanation for $O$ wrt $\<R\cup B,\A,\IC\>$ exists.
\end{enumerate}
\end{theorem}

There are three sources of complexity: negation in programs, recursion
in programs, and the presence of integrity constraints. The next three 
results show that each of these sources by itself pushes the complexity 
up to match the upper bounds of Theorem \ref{thm-upper}.
The first of them concerns the case when the program in an abductive
theory is non-recursive but with negation, and there are no integrity
constraints.

\begin{theorem}\label{thm2}
Let $\A$ be a set of abducible predicates and $R$ a (fixed) non-recursive
program with no abducible predicates in the heads of its rules.
\begin{enumerate}
\item The following problem is coNP-complete: given a set $B$ of abducibles,
an observation $O$, and a pair $(E,F)$ of sets of abducibles, decide whether
$(E,F)$ is a constrained explanation for $O$ wrt $\<R\cup B,\A,\emptyset\>$.
\item The following problem is $\Sigma_2^P$-complete: given a set $B$ of 
abducibles and an observation $O$, decide whether a constrained explanation 
for $O$ wrt $\<R\cup B,\A,\emptyset\>$ exists. 
\end{enumerate}
\end{theorem}

The next theorem shows that when there are no integrity constraints,
disallowing negation, that is, restricting attention to Horn programs,
also leads to the same complexity results as long as we allow recursion.

\begin{theorem}
\label{thm4}
Let $\A$ be a set of abducible predicates and $R$ a (fixed) Horn
program with no abducible predicates in the heads of its rules. 
\begin{enumerate}
\item The following problem is coNP-complete: given a set $B$ of abducibles,
an observation $O$, and a pair $(E,F)$ of sets of abducibles, decide whether
$(E,F)$ is a constrained explanation for $O$ wrt $\<R\cup B,\A,\emptyset\>$.
\item The following problem is $\Sigma_2^P$-complete: given a set $B$ of abducibles and an observation $O$, decide whether a constrained explanation  
for $O$ wrt $\<R\cup B,\A,\emptyset\>$ exists. 
\end{enumerate}
\end{theorem}

The third result addresses the case of abductive theories
with particularly simple programs, namely, non-recursive Horn, but with
integrity constraints.

\begin{theorem}
\label{thmA}
Let $\A$ be a set of abducible predicates, $R$ a (fixed) non-recursive
Horn program with no abducible predicates in the heads of its rules, and
$\IC$ a (fixed) set of integrity constraints.
\begin{enumerate}
\item The following problem is coNP-complete: given a set $B$ of abducibles,
an observation $O$, and a pair $(E,F)$ of sets of abducibles, decide whether
$(E,F)$ is a constrained explanation for $O$ wrt $\<R\cup B,\A,\IC\>$.
\item The following problem is $\Sigma_2^P$-complete: given a set $B$ of abducibles and an observation $O$, decide whether a constrained explanation
for $O$ wrt $\<R\cup B,\A,\IC\>$ exists.
\end{enumerate}
\end{theorem}

On the other end of the spectrum, we have a particularly simple case
when neither of the three sources of complexity is present: the case
of abductive theories with non-recursive Horn programs and without
integrity constraints. For this class of theories the two problems are
tractable.

\begin{theorem}
\label{thm1}
Let $\A$ be a set of abducible predicates and $R$ a (fixed) non-recursive
Horn program with no abducible predicates in the heads of its rules.
The following problems are in P.
\begin{enumerate}
\item Given a set $B$ of abducibles, an observation $O$, and a pair 
$(E,F)$ of sets of abducibles, decide whether $(E,F)$ is a constrained 
explanation for $O$ wrt $\<R\cup B,\A,\emptyset\>$.
\item Given a set $B$ of abducibles and an observation $O$, decide whether 
a constrained explanation for $O$ wrt $\<R\cup B,\A,\emptyset\>$ exists.
\end{enumerate}
\end{theorem}

\section{Related Work}\label{Related_Work}

Abduction was introduced to artificial intelligence in early 1970s by
Harry Pople Jr.  \shortcite{Pople73}, where it is now commonly understood as
the \emph{inference to the best explanation} \cite{Josephson94}.
Over the years several criteria have been proposed to identify the preferred
(best) explanations, all rooted in the Occam's razor parsimony principle.
The most commonly considered one is
subset-minimality \cite{Bylander1991,Konolige1992,Selman-Levesque1990}. A
more restrictive condition of minimum cardinality has also been broadly
studied \cite{Peng-Reggia1990}.
The abduction reasoning formalism we study in the paper uses logic programs
to represent background knowledge in abductive theories. It is referred to
as \emph{abductive logic programming} 
\cite{Eshghi-Kowalski1989,KakasKT92,Dung1991}.
Abductive explanations which allow the removal of hypotheses are first introduced by
Inoue and Sakama \shortcite{Inoue95abductiveframework}.
The importance of abductive
logic programming to knowledge representation was argued by Denecker
and Schreye \shortcite{Denecker1995}. It was applied in diagnosis
\cite{Console*1996}, planning \cite{Eshghi1988,PereiraJ*2004}, natural
language understanding \cite{Balsa97dataloggrammars}, and case-based
reasoning \cite{Satoh1997}. Denecker and Kakas \shortcite{Denecker-Kakas2002}
provide a comprehensive survey of the area.

The complexity of abductive reasoning was first studied by Bylander et al.
\shortcite{Bylander1991}. Eiter et al. \shortcite{EiterGL97}
studied the complexity of reasoning tasks in the abductive logic
programming setting.

The profusion of abductive explanations was explicitly noted by Maher 
\shortcite{mah05} in his work on constraint abduction. Maher
considers a differennt setting and handles the problem by different 
techniques. The key similarity is that in Maher's setting, as in ours,
symbols from the vocabulary not present in the theory can give rise to 
alternative explanations.

None of the earlier works on abduction considered the concepts of
constrainedness or arbitrariness. These concepts were proposed by us
for the setting of view updates in deductive databases \cite{Car*2012}.
View updating consists of modifying base relations to impose properties
on view relations, that is, relations defined on the database by queries.
The subject has received much attention in database research (cf.
the survey papers by Fraternali and Paraboschi
\shortcite{DBLP:conf/rules/FraternaliP93}, and Mayol and Teniente
\shortcite{DBLP:conf/er/MayolT99}).
There is a natural connection between view updating and abduction
\cite{KakasM90,Console:1995:RAD:202787.202795}. The view plays the role
of the background theory, all base relation symbols are abducible predicates,
and
requests for a view update are observations. The role of integrity
constraints is the same in both areas. However, there is an important
distinction here. In view updating, integrity constraints concern only
base relation, a restriction not present in abductive settings.
Our present work adapts the notions of constrainedness and arbitrariness
to the more general setting of abduction. Importantly, we introduce the
new concept of the degree of arbitrariness, which allows us to compare
explanations even when no constrained explanations exist.

\section{Concluding Remarks}
\label{Conclusions}

We proposed a new approach to limiting the space of explanations in
abductive logic programming extended by integrity constraints. Specifically,
we introduced the degree of arbitrariness as a measure of the
quality of an explanation. It imposes a hierarchy on the space of explanations
(possibly already narrowed down by means of other principles): the lower the
degree of arbitrariness, the more compelling an explanation. Explanations
with the degree of arbitrariness equal to 0 are particularly important.
We presented a detailed account of the complexity of reasoning with
constrained explanations when programs in abductive theories are stratified.
In our discussion in Sections 2 and 3 we were implicitly assuming that
the set $\sem_S(P)$ consisted of two-valued interpretations. However,
our definitions 
can also be extended to the three-valued well-founded semantics.
Moreover, since for stratified programs the
well-founded and the stable-model semantics coincide, 
all complexity
results we obtained hold for that case, too. Finally, our approach 
applies to each of the four basic models of abductive reasoning, three of 
which have been studied before, with the remaining one (Definition \ref{def3})
to the best of our knowledge being new. 
Our discussion and the results
suggest that the notions of the degree of arbitrariness and constrainedness
are important additions to the space of fundamental principles of abductive
reasoning.
We note, however, that as with other principles there are situations
where the minimum degree of arbitrariness may not be the appropriate
principle to use for abduction. For example, let us consider the
transitive closure program containing two rules:
$
tc(x,y) \leftarrow \ r(x,y) \mbox{ and }
tc(x,y) \leftarrow \ r(x,z), tc(z,y)
$
with an abducible predicate $r$. If we observe $tc(a,b)$, the explanation
$E_1=(\{r(a,b)\},\emptyset)$ is constrained while $E_2=(\{r(a,z),r(z,b)\},\emptyset)\}$
is arbitrary ($z$ can be replaced by any constant in the vocabulary of the
language). Thus, according to the minimum degree of arbitrariness criterion,
$E_1$ is preferred to $E_2$. However, if the relation $tc$ is the ancestor
relation, $r$ is the parent relation and $ancestor(a,b)$ is observed,
preferring the constrained explanation $\{parent(a,b)\}$ to $\{parent(a,c),
parent(c,b)\}$, which is arbitrary, may be a point of dispute. Establishing
conditions to help decide which
criteria to use when is a grand challenge of the area of abduction. For
the example we just gave, we note that the minimum cardinality
principle would give the same result as the minimum arbitrariness one and
would be open to the same question.
Our work opens several avenues for future research. First, we do not
have a clear picture of the complexity of the case of abductive
theories with non-recursive Horn programs and with integrity
constraints restricted only to abducible predicates. This case is of
interest due to its view updating applications. Next, there is a
challenging problem of resolving the complexity of reasoning with
constrained explanations in the case of abductive theories with
arbitrary programs. 

\section{Acknowledgments}
The third author was supported by the NSF grant IIS-0913459.


\newpage
\section*{Appendix -- Proofs}

\medskip
\noindent

\noindent
\textit{Theorem 1}\\
Let $\T=\<\PP,\A,\IC\>$ be an abductive theory and $(E,F)$ a constrained
explanation of an observation $O$. Then every constant symbol occurring
in $(E,F)$ occurs in $\T$ or in $O$.

\noindent
\begin{proof}
Since $F\subseteq \PP$ (as required by the definition of an explanation),
every constant occurring in $F$ occurs in $\PP$. If $\alpha$ is a constant
appearing in $E$ but not in $\T$ nor in $O$, then changing $\alpha$ to a
fresh constant $\xi$ results in an explanation\footnote{We tacitly
assume here that the semantics of logic programs we consider here are
insensitive to the renaming of constants. All standard semantics of
programs have this property.} and so, contradicts the
constrainedness of $(E,F)$.
\end{proof}

\vspace{0.5mm}
\noindent
\textit{Theorem 2}\\
For every abductive theory $\T=\<\PP,\A,\IC\>$, where $\PP$ is stratified and
interpreted under the stable-model semantics,
and for every observation $O$,
\small
$$ \Psi_{A} (O,\T)= \Psi_{B} (O,\T) =
\Psi_{C} (O,\T) = \Psi_{D} (O,\T).$$
\normalsize

\noindent
\begin{proof}
The assertion follows by the fact that a stratified program admits
exactly one stable model.
\end{proof}

\vspace{0.5mm}
\noindent
\textit{Theorem 3}\\
Let $\A$ be a set of abducible predicates, $R$ a (fixed) stratified
program with no abducible predicates in the heads of its rules, and
$\IC$ a (fixed) set of integrity constraints.
\begin{enumerate}
\item The following problem is in coNP: given a set $B$ of abducibles,
an observation $O$, and a pair $\Delta=(E,F)$ of sets of abducibles,
decide whether $\Delta$ is a constrained explanation for $O$ wrt $\<R\cup B,\A,\IC\>$.
\item The following problem is in $\Sigma_2^P$: given a set $B$ of
abducibles and an observation $O$, decide whether a constrained
explanation for $O$ wrt $\<R\cup B,\A,\IC\>$ exists.
\end{enumerate}
\begin{proof}
(1) The complementary problem consists of deciding that $(E,F)$ is not
an explanation or that is an arbitrary explanation. The following
non-deterministic polynomial-time algorithm decides this problem. Since
$R$ is stratified, one can compute the only stable model,
say $M$, of $R\cup ((B\cup E) \setminus F)$ in time linear in the size of
$B$ and $(E,F)$. If $E$ and $F$ are not disjoint (which can be checked
efficiently), or if $M\not\models O$, or if $M\not\models \IC$, the
$(E,F)$ is not an explanation. Otherwise, $(E,F)$ is an explanation
and we proceed as follows. We non-deterministically guess the set $C$ of
occurrences of some constant $c$ occurring in $E$. We then compute
$(E',F)$ by replacing all occurrences of $c$ mentioned in $C$ with a
fresh constant $\xi$ and, in the same way as before, determine whether
$(E',F)$ is an explanation of $O$.

\smallskip
\noindent
(2) If $(E,F)$ is a constrained explanation, then $E$ and $F$ consist
of abducibles involving only constants appearing in $\T$ and $O$
(cf. Theorem \ref{relevant}).
It follows, that if $(E,F)$ is a constrained explanation, the size of
$E\cup F$ is polynomial in the size of the input. Thus, the problem can
be decided by the following non-deterministic polynomial time algorithm
with an oracle: guess sets $E$ and $F$ of abducibles and check that $(E,F)$
is a constrained explanation. By (1), that task can be decided by a call to
a coNP oracle.
\end{proof}

\noindent
\textit{Theorem 4}\\
Let $\A$ be a set of abducible predicates and $R$ a (fixed) non-recursive
program with no abducible predicates in the heads of its rules.
\begin{enumerate}
\item The following problem is coNP-complete: given a set $B$ of abducibles,
an observation $O$, and a pair $(E,F)$ of sets of abducibles, decide whether
$(E,F)$ is a constrained explanation for $O$ wrt $\<R\cup B,\A,\emptyset\>$.
\item The following problem is $\Sigma_2^P$-complete: given a set $B$ of
abducibles and an observation $O$, decide whether a constrained explanation
for $O$ wrt $\<R\cup B,\A,\emptyset\>$ exists.
\end{enumerate}

\begin{proof}
(1)
The membership part was established in Theorem \ref{thm-upper}. Thus, it
suffices to show the hardness part.

We note that the following version of the SAT problem is NP-complete
(membership is evident, hardness follows by a straightforward reduction
from SAT):
\begin{description}
\item[Input:] A set of atoms $Y$ and a CNF formula $F$ over $Y$ that is
not satisfied by the all-false assignment
\item[Question:] Is $F$ satisfiable (is the QBF formula $\exists Y\,F$ true)?
\end{description}
We will reduce that problem to the problem whether (under the notation
in the statement of the theorem) an explanation $(E,F)$ is arbitrary.

Let then $Y$ be a set of atoms and $F$ a CNF theory that is not satisfied
by the all-false assignment on $Y$. We denote by $\Cl(F)$ the set of clauses
in $F$. Let us consider the vocabulary $\sigma$ consisting of predicate
symbols $bad/0$, $in_Y/1$, $\clause/1$, $\pos/2$, $\ngtd/2$, $\chse/2$,
$\gate/1$, $\true/1$, $\holds/1$, $\sometrue/0$, $\allfalse/0$, $\sat/0$,
$\clfalse/0$ and $\goal/0$, and an abductive theory
$$\T(F)=\<T(F),\{\chse\},\emptyset\>,$$
where $T(F)=R\cup B$, $B$ consists of the atoms
\begin{enumerate}
\item $in_Y(a)$, for every $a\in Y$
\item $\gate(0)$, where $0\notin Y$
\item $pos(a,c)$, for every atom $a\in Y$ and clause $c\in\Cl(F)$
such that $a$ occurs non-negated in $c$
\item $ngtd(a,c)$, for every atom $a\in\At(F)$ and clause $c\in\Cl(F)$
such that $a$ occurs negated in $c$
\end{enumerate}
and $R$ consists of the rules
\begin{enumerate}
\item $\clause(C) \leftarrow \pos(A,C)$
\item $\clause(C) \leftarrow \ngtd(A,C)$
\item $\true(A) \leftarrow in_Y(A), \gate(W), \nf \chse(A,W)$
\item $\holds(C) \leftarrow \pos(A,C), \true(A)$
\item $\holds(C) \leftarrow \ngtd(A,C), \nf \true(A)$
\item $\clfalse \leftarrow \clause(C), \nf \holds(C)$
\item $\sat \leftarrow \nf \clfalse$
\item $\sometrue \leftarrow in_Y(A), true(A)$
\item $\allfalse \leftarrow \nf sometrue$
\item $bad \leftarrow \chse(A,W), \nf in_Y(A)$
\item $\goal \leftarrow \allfalse, \nf bad$
\item $\goal \leftarrow \sat, \nf bad.$
\end{enumerate}
Let $\{\goal\}$ be the set of observed atoms. It is clear that
$U=(\{\chse(a,0)\colon a\in Y\},\emptyset)$ is an explanation ($\goal$ is
derived through the first of its two rules). If $F$ is satisfiable, then let
$Y'\subseteq Y$ be (the representation of) an assignment that satisfies $F$.
One can check that $(\{\chse(a,0)\colon a\in( Y\setminus Y')\}\cup\{\chse(a,\xi)\colon
a\in Y'\},\emptyset)$ is an explanation (now, $\goal$ can be derived via its
second rule). Moreover, $Y'\not=\emptyset$ (by our restriction on the
class of formulas). Thus, $(\{\chse(a,0)\colon a\in Y\},\emptyset)$ is
arbitrary.

Conversely, let us assume that $U=(\{\chse(a,0)\colon a\in Y\},\emptyset)$ is
arbitrary. Then replacing some occurrences of one of the constants must
yield an explanation. Replacing a constant $a\in Y$ with fresh constant
symbol $\xi$ does not
yield an explanation. Indeed, we would have $\chse(\xi,0)$ and no
$in_Y(\xi)$ in the ``add'' part of the explanation. Thus, $bad$ would
hold and would block any possibility of deriving $goal$. It follows that
one or more occurrences of 0 can be replaced by $\xi$ so that the result,
$(\{\chse(a,0)\colon a\in( Y\setminus Y')\}\cup\{\chse(a,\xi)\colon
a\in Y'\},\emptyset)$, is an explanation of $\goal$. Here $Y'\subseteq Y$
is the \emph{non-empty} set of elements in $Y$ identifying the occurrences
of $0$ replaced by $\xi$. Since $Y'\not=\emptyset$,
$\goal$ is derived via the second rule. It follows that $\sat$ is derivable
and so, every clause in $F$ \emph{holds} in the interpretation that assigns
true to all elements of $Y'$ and false to all other elements of $Y$.
Thus, $F$ is satisfiable.

It follows that deciding whether an explanation $(E,F)$ is arbitrary
is NP-hard. Since every explanation is either arbitrary or constrained,
the problem to decide whether $(E,F)$ is constrained is coNP-hard.

\smallskip
\noindent
(2)
As before, the membership part of the assertion follows from Theorem
\ref{thm-upper}.
To prove the hardness part, we note that the following problem is
$\Sigma_2^P$-complete:
\begin{description}
\item[Input:] Two disjoint sets $X$ and $Y$ of atoms, and a DNF formula
$G$ over $X\cup Y$ such that for every truth assignment $v_X$ to atoms
in $X$, the \emph{all-false} assignment to atoms in $Y$ is a model of
formula $G|_{v_X}$
\item[Question:] Is the quantified boolean formula
$\Phi=\exists X \forall Y\, G$ true.
\end{description}
We will reduce it to our problem.

Let $F$ be the CNF obtained from $\neg G$ by applying the De Morgan's
and the double negation laws. Clearly, $F\equiv \neg G$. Let $Cl(F)$ be
the set of clauses of $F$. Let us consider the vocabulary $\sigma$ consisting
of predicate symbols $in_X$, $in_Y$, $\clause/1$, $\pos/2$,
$\ngtd/2$, $\chse/2$, $\gate/1$, $\true_X/1$, $\true_Y/1$, $\true/1$, $\holds/1$,
$\sometrue/0$, $\allfalse/0$, $\sat/0$, $\clfalse/0$, $bad/0$, $good/1$,
and $\goal/0$, and an abductive theory
$$\T(F)=\<T(F),\{\true_X,\chse\},\emptyset\>,$$
where $T(F)=R\cup B$, $B$ consists of the atoms:
\begin{enumerate}
\item $in_X(a)$, for every $a\in X$
\item $in_Y(a)$, for every $a\in Y$
\item $\gate(0)$, where $0\notin Y$
\item $pos(a,c)$, for every atom $a\in X\cup Y$ and clause $c\in\Cl(F)$
such that $a$ occurs non-negated in $c$
\item $ngtd(a,c)$, for every atom $a\in X\cup Y$ and clause $c\in\Cl(F)$
such that $a$ occurs negated in $c$
\end{enumerate}
and $R$ consists of the rules
\begin{enumerate}
\item $\clause(C) \leftarrow \pos(A,C)$
\item $\clause(C) \leftarrow \ngtd(A,C)$
\item $\true_Y(A) \leftarrow in_Y(A), \gate(W), \nf \chse(A,W)$
\item $\true(A) \leftarrow \true_X(A)$
\item $\true(A) \leftarrow \true_Y(A)$
\item $\holds(C) \leftarrow \pos(A,C), \true(A)$
\item $\holds(C) \leftarrow \ngtd(A,C), \nf \true(A)$
\item $\clfalse \leftarrow \clause(C), \nf \holds(C)$
\item $\sat \leftarrow \nf \clfalse$
\item $\sometrue \leftarrow in_Y(A), true_Y(A)$
\item $\allfalse \leftarrow \nf sometrue$
\item $bad \leftarrow \chse(A,W), \nf in_Y(A)$
\item $bad \leftarrow \true_X(A), \nf in_X(A)$
\item $good(A) \leftarrow in_Y(A), \chse(A,W)$
\item $bad \leftarrow in_Y(A), \nf good(A)$
\item $\goal \leftarrow \allfalse, \nf bad$
\item $\goal \leftarrow \sat, \nf bad$.
\end{enumerate}
Let $O=\{\goal\}$ be an observation. We will prove that $\Phi$ is true if and
only if $\goal$ has a constrained explanation from $\T(F)$.\\
($\Rightarrow)$ Let $v_X$ be an assignment of truth values to variables in
$X$ such that the formula $\forall Y\,G|_{v_X}$ is true. Here by $G|_{v_X}$
we denote the formula obtained from $G$ by substituting the truth values
given by $v_X$ for the corresponding variables from $X$, and then by simplifying
these values away. We understand the formula $F|_{v_X}$ in the same way.
Clearly, $F|_{v_X} \equiv \neg G|_{v_X}$. Thus, $\exists Y F|_{v_X}$
is false.\\
Let us define
$$E=\{\true_X(a): \mbox{$a\in X$ and $v_X(a)=\true$}\}\cup$$
$$\{\chse(a,0)\colon a\in Y\}.$$
We will show that $(E,\emptyset)$ is a constrained explanation of $\goal$.
First, it is evident that $(E,\emptyset)$ is an explanation as $\goal$ can
be derived through the first of its two rules. Next, we note that we cannot
replace any constant $a$ appearing in atoms $\true_X(a)$ with a new constant
$\xi$. Indeed, if $\true_X(\xi)$ were to be a part of a the ``add'' part of
an explanation, $\bad$ would hold (via the rule (13)) and $\goal$ would
not!
Similarly, we cannot replace $a\in Y$ in any atom $\chse(a,0)$, as
only elements of $Y$ must show on these positions, the property forced
by rule (12).
Finally, we cannot replace any non-empty set of 0's with
$\xi$. If any such replacement resulted in an explanation, $\goal$ could
only be derived through its second clause ($\allfalse$ cannot be derived
now). However, that would imply that $\exists Y F|_{v_X}$ is true, with
the ``witness'' assignment assigning \emph{true} to every $y\in Y$
such that $\chse(y,\xi)$ is a part of the modified explanation, and
\emph{false} to all other elements of $Y$.

\smallskip
\noindent
($\Leftarrow$) Let us assume that $goal$ has a constrained explanation.
It must have a form $(E,\emptyset)$, where
$$E=\{\true_X(a): \mbox{$a\in U$}\}\cup\{\chse(a,b)\colon
a\in Y, b\in U_a\},$$
where $U$ is some subset of $X$ and where for every $a\in Y$, $U_a$,
is some \emph{nonempty} set. Indeed, if for some $a\in Y$, there is no
$b$ such that $\chse(a,b)\in E$, $good(a)$ cannot be derived from the
revised program and, consequently, $\bad$ would follow (by the rule (15)).
That would make it impossible to derive $\goal$.

Now, if for some $a\in Y$ there is $b\in U_a$ such that $b\not=0$,
then $(E,\emptyset)$ is not constrained (indeed, that constant $b$
could be replaced by a new constant $\xi$ without any effect on
the derivability of $goal$). Thus, for every $a\in Y$, $U_a=\{0\}$
and so,
$$E=\{\true_X(a): \mbox{$a\in U$}\}\cup\{\chse(a,0)\colon
a\in Y\}.$$
Since this explanation is constrained, there is no subset of positions
where 0 occurs that can be substituted with $\xi$. Therefore, $\exists Y
F|_{v_X}$, where $v_X$ is the truth assignment determined by $U$, is false.
One can show that by following the argument used in part (1) of the theorem
(due to our assumption on $G$, the all-false assignment to atoms in $Y$ is
not a model of $F|_{v_X}$). Thus, $\Phi$ is true.~\end{proof}

\noindent
\textit{Theorem 5}\\
Let $\A$ be a set of abducible predicates and $R$ a (fixed) Horn
program with no abducible predicates in the heads of its rules.
\begin{enumerate}
\item The following problem is coNP-complete: given a set $B$ of abducibles,
an observation $O$, and a pair $(E,F)$ of sets of abducibles, decide whether
$(E,F)$ is a constrained explanation for $O$ wrt $\<R\cup B,\A,\emptyset\>$.
\item The following problem is $\Sigma_2^P$-complete: given a set $B$ of abducibles and an observation $O$, decide whether a constrained explanation
for $O$ wrt $\<R\cup B,\A,\emptyset\>$ exists.
\end{enumerate}

\noindent
\begin{proof}
(1)
The membership part follows by Theorem \ref{thm-upper}. To prove hardness,
we show that the problem to decide whether $(E,F)$ is arbitrary is NP-hard.
That is sufficient, as every explanation is either arbitrary or constrained.
To show NP-hardness of the problem to decide whether an explanation is
arbitrary, we reduce the SAT problem to it.

Thus, let $Y$ be a (finite) set of atoms, say $Y=\{y_1,\ldots,y_n\}$,
and $F$ a CNF consisting of clauses $c_1,\ldots, c_m$. We denote by $\Cl(F)$
the set of clauses in $F$, that is, $\Cl(F)=\{c_1,\ldots, c_m\}$. Let us
also consider three additional distinct symbols $t$, $f$ and $0$. We define
the vocabulary $\sigma$ to consist of predicate symbols
$\inn_Y/1$, $\clause/1$, $\pos/2$, $\ngtd/2$, $p/2$
$\true/1$, $\false/0$, $\ok/1$, $\next/2$, $\next_C/2$,
$\clsat/0$, $\sat/1$, and $\goal/0$, and an abductive theory
$$\T(F)=\<T(F),\{p\},\emptyset\>,$$
where $T(F)$ consists of the following atoms:
\begin{enumerate}
\item $\inn_Y(a)$, for every $a\in Y\cup\{t,f\}$
\item $\pos(a,c)$, for every atom $a\in Y$ and clause $c\in\Cl(F)$
such that $a$ occurs non-negated in $c$
\item $\ngtd(a,c)$, for every atom $a\in Y$ and clause $c\in\Cl(F)$
such that $a$ occurs negated in $c$
\item $\next(y_i,y_{i+1})$, for $i=1,\ldots, n-1$, $\next(t,y_1)$, and
$\next(y_n,f)$
\item $\next_C(c_i,c_{i+1})$, for $i=1,\ldots, m-1$
\item $p(t,0)$
\end{enumerate}
and of the following rules
\begin{enumerate}
\item $\clause(C) \leftarrow \pos(A,C)$
\item $\clause(C) \leftarrow \ngtd(A,C)$
\item $\true(A) \leftarrow \inn_Y(A), p(A,Z), p(t, Z)$
\item $\false(A) \leftarrow \inn_Y(A), p(A,Z), p(f, Z)$
\item $\clsat(C) \leftarrow \pos(A,C), \true(A)$
\item $\clsat(C) \leftarrow \ngtd(A,C), \false(A)$
\item $\ok(t)$
\item $\ok(A) \leftarrow \ok(A'), \next(A',A),\true(A)$
\item $\ok(A) \leftarrow \ok(A'), \next(A',A),\false(A)$
\item $\ok(f) \leftarrow \ok(A'), \next(A',f)$
\item $\sat(c_1) \leftarrow \clsat(c_1)$
\item $\sat(C) \leftarrow \sat(C'), \next_C(C',C), \clsat(C)$
\item $goal \leftarrow \ok(f), \sat(c_m), p(f,Z)$.
\end{enumerate}

Clearly, the pair $(E,\emptyset)$, where $E=\{p(x,0)\colon x\in Y\cup
\{f\}\}$, is an explanation of $goal$. Indeed, for every $x\in Y$, both
$\true(x)$ and $\false(x)$ can be derived from $T(F)\cup E$ (because
$p(t,0)$ and $p(f,0)$ both hold in $T(F)\cup E$). Thus, for every clause
$c$, $\clsat(C)$ can be derived, too. These two observations imply that
$\ok(f)$ and $\sat(c_m)$ can both be derived from $T(F)\cup E$.
Consequently, $\goal$ is explained by $(E,\emptyset)$.

Let us assume that $E$ is arbitrary. We will prove that $F$ is satisfiable.
By the definition, one of the constants
appearing in $E$ can be replaced by a fresh constant $\xi$ so that the
resulting pair $(E',\emptyset)$ is an explanation of $\goal$ wrt $\T(F)$.
It follows that $\ok(f)$ can be derived from $T(F)\cup E'$, that is,
that for every $x\in Y$, at least one of $\true(x)$ and $\false(x)$
can be derived. This, implies that for every $x\in Y$, $p(x,0)\in T(F)
\cup E'$, that is, $\xi$ is substituted for $f$ or 0 in $E$.

Since, by (13), every explanation of $goal$ contains at least one atom of
the form $p(f,z)$, $\xi$ is not substituted for $f$ in $E$ to produce $E'$.
Thus, $E'$
is obtained from $E$ by substituting $\xi$ for some occurrences of 0.
Let $U= \{u\in Y\cup\{f\}\colon p(u,\xi)\in E'\}$. If $f\notin U$, then
let $y$ denote any element in $U\cap Y$ (such an element exists as
$U\not= \emptyset$). Since $p(t,\xi)$ and $p(f,\xi)$ are not in $T(F)
\cup E'$, neither $\true(y)$ nor $\false(y)$ can be derived from $T(F)
\cup E'$. Thus, neither $\ok(f)$ nor $goal$ can be derived from $T(F)
\cup E'$. It follows that $f\in U$. Consequently, for every $x\in U$,
$\false(x)$ can be derived from $T(F)\cup E'$, and $\true(x)$ cannot be.
Similarly, for every $x\in Y\setminus U$, $\true(x)$ can be derived
from $T(F)\cup E'$, and $\false(x)$ cannot be. Thus, the atoms $\true(x)$
and $\false(x)$ in $T(F)\cup E'$ determine a truth assignment on atoms
of $Y$. Since $\sat(c_m)$ can be derived from $T(F)\cup E'$, $\clsat(c)$
can be derived form $T(F)\cup E'$, for every clause $c$ in $F$. It follows
that the truth assignment determined by the atoms $\true(x)$ and
$\false(x)$ in $T(F)\cup E'$ satisfies $F$.

Conversely, let us assume that $F$ is satisfiable. Let us consider any
satisfying assignment for $F$ and let $U$ comprises $f$ and those atoms
in $Y$ that are false under this assignment. Let $E'$ be obtained from
$E$ by substituting $\xi$ for the occurrences of 0 in atoms $p(y,0)$, $y
\in U$. It is easy to verify that for every $y\in U$, $\false(y)$ can be
derived from $T(F)\cup E'$, and $\true(y)$ cannot be. Similarly, for
every $y\in Y\setminus U$, $\true(y)$ can be derived from $T(F)\cup E'$,
and $\false(y)$ cannot be. Moreover, $\clsat(c)$ can be derived from
$T(F)\cup E'$, for every clause $c$ of $F$. Consequently, $\ok(f)$
and $\sat(c_m)$ can be derived from $T(F)\cup E'$. Since $p(f,\xi)\in E'$,
$goal$ can be derived from $T(F)\cup E'$, that is, $(E',\emptyset)$ is
an explanation of $\goal$ wrt $\T$. Thus, $(E,\emptyset)$ is arbitrary.

\smallskip
\noindent
(2)
The argument for the membership part follows by
Theorem \ref{thm-upper}.

We prove hardness.
The problem to decide whether a QBF $\Phi=\exists X\forall Y G$,
where $G$ is a DNF formula over variables in $X\cup Y$, is true, is
$\Sigma_2^P$-complete. We will reduce it to the problem in question.

Below, we understand $v_X$, $F$, $\Cl(F)$ and $G|_{v_X}$
as in the proof of Theorem \ref{thm2}. 
We assume that $X=\{x_1,\ldots,
x_k\}$, $Y=\{y_1,\ldots,y_n\}$ and $C=\{c_1,\ldots,c_m\}$.

We define $\sigma$ to consist of predicate symbols $\inn_X/1$, $\inn_Y/1$,
$\clause/1$, $\pos/2$,
$\ngtd/2$, $\next_X/2$, $\next_Y/2$, $next_C/2$, $\true_X/1$, $\false_X/1$,
$\true/1$, $\false/1$, $\ok_X/1$, $\ok_Y/1$, $\sat/1$, $f_X/1$, $l_X/1$, $f_Y/1$,
$l_Y/1$, $f_C/1$, $l_C/1$, $\tr/1$, $\fa/1$, $\assign/2$,
$\good_X/0$, $\good_Y/0$, $\good_C/0$, $\good_f/0$, $\goal/0$. We assume
three new distinct constants $0$, $t$ and $f$ and consider an abductive theory
$$\T(F)=\<T(F),\{\true_X,\;\false_X,\;\assign,\; \fa\},\emptyset\>,$$
where $T(F)$ consists of the following atoms (part $B$):
\begin{enumerate}
\item $\inn_X(a)$, for every $a\in X$
\item $\inn_Y(a)$, for every $a\in Y\cup\{t,f\}$
\item $\pos(a,c)$, for every atom $a\in X\cup Y$ and clause $c\in\Cl(F)$
such that $a$ occurs non-negated in $c$
\item $\ngtd(a,c)$, for every atom $a\in X\cup Y$ and clause $c\in\Cl(F)$
such that $a$ occurs negated in $c$
\item $f_X(x_1)$, $l_X(x_k)$
\item $f_Y(y_1)$, $l_Y(y_n)$
\item $f_C(c_1)$, $l_C(c_m)$
\item $\next_X(x_i,x_{i+1})$, for $i=1,\ldots, k-1$
\item $\next_Y(y_i,y_{i+1})$, for $i=1,\ldots, n-1$
\item $\next_C(c_i,c_{i+1})$, for $i=1,\ldots, m-1$
\item $tr(0)$
\end{enumerate}
and of the following rules (part $R$)
\begin{enumerate}
\item $\clause(C) \leftarrow \pos(A,C)$
\item $\clause(C) \leftarrow \ngtd(A,C)$
\item $\true(A) \leftarrow \true_X(A)$
\item $\false(A) \leftarrow \false_X(A)$
\item $\true(B) \leftarrow \inn_X(B), \true_X(A),\false_X(A)$
\item $\false(B) \leftarrow \inn_X(B), \true_X(A),\false_X(A)$
\item $\true(B) \leftarrow \inn_Y(B), \true_X(A),\false_X(A)$
\item $\false(B) \leftarrow \inn_Y(B), \true_X(A),\false_X(A)$
\item $\true(A) \leftarrow \inn_Y(A), \assign(A,Z), \tr(Z)$
\item $\false(A) \leftarrow \inn_Y(A), \assign(A,Z), \fa(Z)$
\item $\clsat(C) \leftarrow \pos(A,C), \true(A)$
\item $\clsat(C) \leftarrow \ngtd(A,C), \false(A)$
\item $\ok_X(A) \leftarrow f_X(A), \true(A)$
\item $\ok_X(A) \leftarrow f_X(A), \false(A)$
\item $\ok_X(A) \leftarrow \ok_X(A'), \next_X(A',A),\true(A)$
\item $\ok_X(A) \leftarrow \ok_X(A'), \next_X(A',A),\false(A)$
\item $\good_X \leftarrow  \ok_X(A), \l_X(A)$
\item $\ok_Y(A) \leftarrow f_Y(A), \true(A)$
\item $\ok_Y(A) \leftarrow f_Y(A), \false(A)$
\item $\ok_Y(A) \leftarrow \ok_Y(A'), \next_Y(A',A),\true(A)$
\item $\ok_Y(A) \leftarrow \ok_Y(A'), \next_Y(A',A),\false(A)$
\item $\good_Y \leftarrow  \ok_Y(A), \l_Y(A)$
\item $\sat(C) \leftarrow \clsat(C), f_C(C)$
\item $\sat(C) \leftarrow \sat(C'), \next_C(C',C), \clsat(C)$
\item $\good_C \leftarrow  \sat(C), \l_C(C)$
\item $goal \leftarrow \good_X,\good_Y,\good_C,\fa(Z)$
\item $goal \leftarrow \good_X,\good_Y,\inn_Y(A),\false(A),\true(A),\fa(Z).$
\end{enumerate}

Let $v_X$ be an assignment of truth values to variables in $X$ such that
the formula $\forall Y\;G|_{v_X}$ is true, and let
\begin{align*}
E= &\{\true_X(a)\colon \mbox{$a\in X$ and $v_X(a)=\true$}\} \cup\\
   &\{\false_X(a)\colon \mbox{$a\in X$ and $v_X(a)=\false$}\} \cup\\
   &\{\assign(y,0)\colon y\in Y\} \cup \{\fa(0)\}.
\end{align*}
It is clear that $(E,\emptyset)$ is an explanation for $\goal$ wrt
$\T(F)$. Indeed, since for every $y\in Y$ we have $\true(y)$ and
$\false(y)$, $\goal$ can be derived by means of the rule (27).
Let us assume that $E$ is arbitrary. Then, there is a constant, say $a$,
appearing in $E$ such that replacing some occurrences of $a$ with a fresh 
constant $\xi$ 
results in another explanation of $\goal$. However, if $a\in X$, then 
neither $\true(a)$ nor $\false(a)$ can be derived after the replacement. 
Consequently, we cannot derive $\good_X$ and so, we cannot derive $\goal$ 
either. If $a\in Y$, then again neither $\true(a)$ nor $\false(a)$ can be 
derived. Now, $\good_Y$ cannot be derived and so, neither can $\goal$. 
Thus, $a=0$. If we do not replace the occurrence
of $0$ in $\fa(0)$ with $\xi$, then there is $y\in Y$ such that we replace
the occurrence of $0$ in $\assign(y,0)$ with $\xi$. For that $y$, after
the replacement we cannot derive $\true(y)$ nor $\false(y)$ and so,
$\good_Y$ and $\goal$ cannot be derived. It follows that there is a set 
$Y'\subseteq Y$ such that 
\begin{align*}
E'= &\{\true_X(a)\colon \mbox{$a\in X$ and $v_X(a)=\true$}\} \cup\\
   &\{\false_X(a)\colon \mbox{$a\in X$ and $v_X(a)=\false$}\} \cup\\
   &\{\assign(y,0)\colon y\in Y\setminus Y'\} \cup \{\assign(y,\xi)\colon
       y\in Y'\} \cup \{\fa(\xi)\}
\end{align*}
gives rise to an explanation $(E',\emptyset)$ of $\goal$. Clearly,
after applying $(E',\emptyset)$, for no $y\in Y$, both $\true(y)$ and 
$\false(y)$ can be derived. Thus, 
$\goal$ must be derivable by means of the rule (26). Moreover, for every
$y\in Y$, we have exactly one of $\true(y)$ and $\false(y)$ hold:
$\true(y)$ holds in $y\in Y\setminus Y'$, and $\false(y)$ holds if
$y\in Y'$. Since $\goal$ can be derived, it follows that $\good_C$
can be derived. Consequently, the truth assignment on $Y$ defined
by the atoms $\true(y)$ and $\false(y)$, where $y\in Y$, satisfies 
the set of clauses of $F|_{v_X}$, that is $\exists Y F|_{v_X}$ is true.
This is a contradiction since $\exists Y F|_{v_X} \equiv 
\neg\forall Y\;G|_{v_X}$. Hence, $(E,\emptyset)$ is constrained.

Conversely, let $(E,\emptyset)$ be a constrained explanation of the goal.
Clearly, $E$ consists of facts of the form $\true_X(a)$, $\false_X(b)$, 
$\assign(y,z)$ and $\fa(w)$. For every element $x\in X$, at least one
of $\true_X(x)$ and $\false_X(x)$ must be present in $E$ (otherwise, we 
cannot derive $\good_X$). Moreover, if for at least one element $a\in X$
we have $\true_X(a)$ and $\false_X(a)$ in $E$, then changing these two 
occurrences of $a$ to $\xi$ does not affect derivability of
$\goal$ (indeed, by the rules (5)-(8) both before and after the change 
we have $\true(x)$ and $\false(x)$ hold for all $x\in X\cup Y$). Thus,
$(E,\emptyset)$ would not be constrained. Finally if $\true_X(a)$ or 
$\false_X(a)$ is in $E$, $a\in X$. Otherwise, that $a$ could be replaced 
by $\xi$ without affecting the derivability of $\goal$, contradicting again 
the assumption that $(E,\emptyset)$ is constraied. It follows that
if $\true_X(x)$ or $\false_X(x)$ is in $E$, $x\in X$ and that  
the atoms $\true_X(x)$ and $\false_X(x)$ that belong to $E$ determine
a truth assignment on $X$, say $v_X$. 

Next, let us assume that for some $\alpha\not=0$ we have $\assign(y,\alpha)
\in E$. Then replacing all occurrences of $\alpha$ by $\xi$ (including 
possibly an occurrence if $\alpha$ in $\fa(\alpha)$) has no effect on 
the derivability of $\goal$. As before, we get a contradiction.
Thus, if $E$ contains facts $\assign(y,z)$, they are of the form
$\assign(y,0)$. If any of these $y$'s is not in $Y$, it can be changed
to $\xi$ without affecting the derivability of $\goal$.

Next, we note that if $E$ contains $\fa(\alpha)$, where $\alpha\not=0$, 
that $\alpha$ can be changed to $\xi$ without affecting the derivability 
of $\goal$.

If for some $y\in Y$, $\assign(y,0)$ is not in $E$, then for that
$y$ we can derive neither $\true(y)$ nor $\false(y)$. Thus, we
cannot derive $\good_Y$ and, consequently, we cannot derive $\goal$ 
either. It follows that $E$ contains all facts $\assign(y,0)$, $y\in 
Y$, and no other facts based on the relation symbol $\assign$.

If $\fa(0)$ is not in $E$, $\goal$ cannot be derived. Thus, $E$ is of
the form we considered above. Let $Y'\subseteq Y$ and let $E'$ be as 
above. Since $(E,\emptyset)$ is constrained, $(E',\emptyset)$ is not an 
explanation of $\goal$. That is a truth assignment on $Y$ such that 
elements in $Y\setminus Y'$ are assigned \emph{true} and those in $Y'$
are assigned \emph{false} is not a satisfying assingment for $F|_{v_X}$. 
Consequently, it follows that $\exists Y F|_{v_X}$ is false
and so, $\forall Y G|_{v_X}$ is true. This last property implies that
$\exists X\forall Y G$ is true. 
\end{proof}

\noindent
\textit{Theorem 6}\\
Let $\A$ be a set of abducible predicates, $R$ a (fixed) non-recursive
Horn program with no abducible predicates in the heads of its rules, and
$\IC$ a (fixed) set of integrity constraints.
\begin{enumerate}
\item The following problem is coNP-complete: given a set $B$ of abducibles,
an observation $O$, and a pair $(E,F)$ of sets of abducibles, decide whether
$(E,F)$ is a constrained explanation for $O$ wrt $\<R\cup B,\A,\IC\>$.
\item The following problem is $\Sigma_2^P$-complete: given a set $B$ of abducibles and an observation $O$, decide whether a constrained explanation
for $O$ wrt $\<R\cup B,\A,\IC\>$ exists.
\end{enumerate}

\noindent
\begin{proof}
(1)
The membership part follows by Theorem \ref{thm-upper}. To prove hardness,
we show that the problem to decide whether $(E,F)$ is arbitrary is NP-hard.
That is sufficient, as every explanation is either arbitrary or constrained.
To show NP-hardness of the problem to decide whether an explanation is
arbitrary, we reduce the SAT problem to it. Thus, let $Y$ be a
(finite) set of atoms and $F$ a CNF theory over $Y$. As before, we denote
by $\Cl(F)$ the set of clauses in $F$. Let us also consider three additional
distinct symbols $t$, $f$ and $0$. We define
the vocabulary $\sigma$ to consist of predicate symbols
$\inn_Y/1$, $\clause/1$, $\pos/2$, $\ngtd/2$, $p/2$
$\true/1$, $\false/0$, $\ok/1$, $\next/2$, $\next_C/2$,
$\clsat/0$, $\sat/1$, and $\goal/0$, and an abductive theory
$$\T(F)=\<T(F),\{p\},\IC\>,$$
where $T(F)$ consists of the following atoms:
\begin{enumerate}
\item $\inn_Y(a)$, for every $a\in Y\cup\{t,f\}$
\item $\pos(a,c)$, for every atom $a\in Y$ and clause $c\in\Cl(F)$
such that $a$ occurs non-negated in $c$
\item $\ngtd(a,c)$, for every atom $a\in Y$ and clause $c\in\Cl(F)$
such that $a$ occurs negated in $c$
\item $p(t,0)$
\end{enumerate}
and of the rules
\begin{enumerate}
\item $\clause(C) \leftarrow \pos(A,C)$
\item $\clause(C) \leftarrow \ngtd(A,C)$
\item $\true(A) \leftarrow \inn_Y(A), p(A,Z), p(t, Z)$
\item $\false(A) \leftarrow \inn_Y(A), p(A,Z), p(f, Z)$
\item $\clsat(C) \leftarrow \pos(A,C), \true(A)$
\item $\clsat(C) \leftarrow \ngtd(A,C), \false(A)$
\item $\goal \leftarrow p(f,X)$
\end{enumerate}
and where $\IC$ consists of
\begin{enumerate}
\item $\forall C\; \clause(C) \supset \clsat(C)$
\item $\forall A\; in_Y(A) \supset \false(A)\lor\true(A)$.
\end{enumerate}

Clearly, the pair $(E,\emptyset)$, where $E=\{p(x,0)\colon x\in Y\cup
\{f\}\}$, is an explanation of $goal$. Indeed, for every $x\in Y$, both
$\true(x)$ and $\false(x)$ can be derived from $T(F)\cup E$ (because
$p(t,0)$ and $p(f,0)$ both hold in $T(F)\cup E$). Thus, for every clause
$c$, $\clsat(C)$ can be derived, too. Consequently, the two integrity
constraints in the theory hold for the least model of the program
$T(F)\cup E$. Moreover, $\goal$ belongs to this unique model and so,
it is entailed by the revised theory.

Let us assume that $E$ is arbitrary. We will prove that $F$ is satisfiable.
By the definition of arbitrariness, one of the constants appearing in $E$
can be replaced by a fresh constant $\xi$ so that the resulting pair
$(E',\emptyset)$ is an explanation of $\goal$ wrt $\T(F)$, that is,
the least model of $T(F)\cup E'$ satisfies both integrity constraints of
the abductive theory and contains an atom of the form $p(f,z)$ (in order 
for $\goal$ to hold.

If $f$ is replaced with $\xi$ in $E$, then the least model of $T(F)\cup E'$
does not contain any fact of the form $p(f,X)$, and $(E',\emptyset)$ is
not an explanation.
If some atom $x\in Y$ is replaced by $\xi$, then for that atom neither
$\true(x)$ nor $\false(x)$ belongs to the least model of $T(F)\cup E'$,
which means that this model violates the second integrity constraint,
contrary to the fact that $(E',\emptyset)$ is an explanation.

Thus, there is a non-empty set $U\subseteq Y\cup\{f\}$, such that when each
occurrence of $0$ in $p(x,0)$, where $u\in U$, is replaced by $\xi$,
the resulting set $E'=\{p(x,\xi)\colon x\in U\}\cup \{p(x,0)\colon x\in
(Y\setminus U)\}$ gives rise to an explanation $(E',\emptyset)$. Let us 
assume that $f\notin U$. Since $U\not=\emptyset$, $U\cap Y\neq\emptyset$.
Let $x\in U\cap Y$. For this $x$, the least model of $T(F)\cup E'$ 
contains neither $\false(x)$ nor $\true(x)$, violating the second integrity
constraint. Thus, $f\in U$ and, consequently, the least model of $T(F)\cup 
E'$, contains atoms $\false(x)$, where $x\in U\setminus\{f\}$, and $\true(x)$, 
where $x\in Y\setminus U$. It follows that this set of atoms defines a 
valuation on $Y$. Moreover, since the first integrity constraint holds, 
this valuation satisfies all clauses of $F$.

Conversely, let us assume that $F$ is satisfiable. Let us consider any
satisfying assignment for $F$ and let $U$ comprises $f$ and those atoms
in $Y$ that are false under this assignment. Let $E'$ be obtained from
$E$ by substituting $\xi$ for the occurrences of 0 in atoms $p(y,0)$, $y
\in U$. It is easy to verify that for every $y\in U$, $\false(y)$ can be
derived from $T(F)\cup E'$, and $\true(y)$ cannot be. Similarly, for
every $y\in Y\setminus U$, $\true(y)$ can be derived from $T(F)\cup E'$,
and $\false(y)$ cannot be. Moreover, $\clsat(c)$ can be derived from
$T(F)\cup E'$, for every clause $c$ of $F$. Thus, both integrity constraints
are satisfied by the least model of $T(F)\cup E'$, and that model also contains
$p(f,\xi)$ and so, also $\goal$. Thus, $(E,\emptyset)$ is an arbitrary
explanation of goal.

\smallskip
\noindent
(2)
The membership part follows by Theorem \ref{thm-upper}.
To prove hardness we proceed similarly as in the proofs of Theorems 4 and 5.
\end{proof}

\noindent
\textit{Theorem 7}\\
Let $\A$ be a set of abducible predicates and $R$ a (fixed) non-recursive
Horn program with no abducible predicates in the heads of its rules.
The following problems are in P.
\begin{enumerate}
\item Given a set $B$ of abducibles, an observation $O$, and a pair
$(E,F)$ of sets of abducibles, decide whether $(E,F)$ is a constrained
explanation for $O$ wrt $\<R\cup B,\A,\emptyset\>$.
\item Given a set $B$ of abducibles and an observation $O$, decide whether
a constrained explanation for $O$ wrt $\<R\cup B,\A,\emptyset\>$ exists.
\end{enumerate}

\begin{proof}
(1)
Let us consider an explanation $(E,F)$. Since
$R$ is non-recursive, there is a constant, say $k$, such that any proof
of $o$ based on the rules in $R$ and facts in $B$ revised by $(E,F)$
has length bounded from above by $k$. Thus, the total number of facts
used in any such proof is bounded by $k$, too.

Since at most $k$ atoms in $E$ are relevant to any proof, if $E$
contains more than $k$ abducibles with predicate symbols of positive
arity, it is not constrained. Indeed, at least one of these abducibles
does not play any role in the proof. For for this abducible, say $a =
p(c_1,\ldots,c_m)$, we have that replacing $c_1$ with $\xi$ in $a$ results
in an explanation.

If on the other hand, the number of abducibles with predicate symbols of
positive arity in $E$ is less than or equal to $k$, then the total number
of constants occurring in all abducibles in $E$ is bounded by a constant
$k'$ dependent on $R$ only (independent of the size of input,
that is, of the size of $B$). Thus, there is only a fixed number of
possible selections of occurrences of a constant for replacement by a new
symbol $\xi$. For each of them, we can test in polynomial time whether it
leads to an explanation. Thus, we can decide whether $(E,F)$ is constrained
in polynomial time.

\smallskip
\noindent
(2)
If $(E,F)$ is a constrained explanation, then $(E,\emptyset)$ is
a constrained explanation. Moreover, we can assume that $E$ contains
all zero arity abducibles in $B$. Thus, each such
constrained explanation is determined by its non-zero arity abducibles.
Non-zero arity abducibles in a constrained explanation use only constants
appearing in $O$
and $P$ (Theorem \ref{relevant}).
It follows, the set of all possible non-zero arity abducibles that
might be chosen to form $E$ has size that is polynomial in the size of
$B$ (the input size). Since by an argument from the previous proof, we
can assume that $E$ contains no more than $k$ non-zero arity abducibles
(where $k$ is a constant depending only on $R$), the set of all candidate
explanations that need to be tested to decide the problem is polynomial in
the size of input. Since each such candidate explanation can be tested
for constrainedness in polynomial time (by the previous result), the assertion
follows.
\end{proof}

\label{lastpage}
\end{document}